\documentclass[aps,prx,reprint,twocolumn,notitlepage,superscriptaddress]{revtex4-1}
\usepackage{graphicx}
\usepackage{amsmath,amssymb}
\usepackage{color}
\usepackage{bm}
\usepackage{bbm}
\usepackage{mathtools}

\usepackage{hyperref}
\usepackage[dvipsnames]{xcolor}
\DeclareMathOperator{\arccosh}{\mathrm{arccosh}}

\begin{document}

\title{Rigorous lower bound of the dynamical critical exponent of the Ising model}
\author{Rintaro Masaoka}
\affiliation{Department of Applied Physics, The University of Tokyo, Tokyo 113-8656, Japan}
\author{Tomohiro Soejima
}
\affiliation{Department of Physics, Harvard University, Cambridge, MA 02138, USA}
\author{Haruki Watanabe}\email{hwatanabe@g.ecc.u-tokyo.ac.jp}
\affiliation{Department of Applied Physics, The University of Tokyo, Tokyo 113-8656, Japan}
\date{\today}

\begin{abstract}
We study the kinetic Ising model under Glauber dynamics and establish an upper bound on the spectral gap for finite systems.
This bound implies the critical exponent inequality $z \geq 2$, thereby rigorously improving the previously known estimate $z \geq 2 - \eta$.
Our proof relies on the mapping from stochastic processes to frustration-free quantum systems and leverages the Simon--Lieb and Gosset--Huang inequalities.
\end{abstract}

\maketitle

\section{Introduction}
The Ising model is a paradigmatic model of phase transition. As the namesake model for the Ising universality class, its critical exponents in particular have been studied extensively with numerical methods~\cite{hasenbuschCriticalExponentsThreedimensional1999,Poland}.
Rigorous results, including the exact solution in two dimensions, have also been explored~\cite{onsagerCrystalStatisticsTwoDimensional1944,fernandez}.
An important version of the Ising model is its \textit{stochastic} variant, which captures how spins equilibrate when there are no conservation laws. 
The kinetic Ising model, which implements the Glauber dynamics of the Ising model, belongs to this universality class.

When the Ising model is near or right at criticality, the relaxation time $\tau$ is believed to undergo critical slowing down.
The degree of the critical slowing down can be captured by the dynamical critical exponent. 
Near the critical point in an infinite system, $\tau$ is expected to behave as
\begin{equation}
\tau \simeq \xi^z,\label{defz1}
\end{equation}
where $\xi$ is the correlation length, which also is expected to grow as $\xi \simeq |T-T_c|^{-\nu}$. For the kinetic Ising model, Halperin~\cite{PhysRevB.8.4437,10.1143/PTP.41.941} showed $z\nu \geq \gamma$, where $\gamma$ is the critical exponent of the susceptibility i.e. $\chi \simeq |T-T_c|^{-\gamma}$~\footnote{We review the discussion in Ref.~\cite{PhysRevB.8.4437} in Appendix~\ref{sec:Halperin}.}.
Using (unproven) Fisher's identity $\gamma=(2-\eta)\nu$, this bound is converted to
\begin{align}
z \geq 2-\eta,\label{previous}
\end{align}
where $\eta$ is the anomalous dimension characterizing the decay of correlation functions at the critical point, i.e., $\langle \sigma_i \sigma_j \rangle \simeq |i - j|^{-d+2-\eta}$ as $|i-j|\to \infty$. The values of $\eta$ are known exactly in $d=2$, and $d\geq 4$, and very accurately in $d=3$; $\eta=1/4$ for $d=2$, $\eta=0.0362978(20)$ for $d=3$~\cite{Poland}, and $\eta=0$ for $d\geq4$~\cite{fernandez}.  Numerical results performed with models in the stochastic Ising model give $z=2.1667(5)$ for $d=2$~\cite{PhysRevB.62.1089} and $z=2.0245(15)$ for $d=3$~\cite{PhysRevE.101.022126}, which are consistent with this bound. A recent five-loop $\epsilon$-expansion analysis yields compatible estimates, $z = 2.14(2)$ for $d = 2$ and $z = 2.0236(8)$ for $d = 3$ \cite{adzhemyanDynamicCriticalExponent2022}.

Equivalently, right at the critical point in a finite system, the system size $L$ replaces the correlation length: 
\begin{equation}
\tau \simeq L^z.\label{defz2}
\end{equation}
The best rigorous bound so far, based on this definition, for the two-dimensional kinetic Ising model is $z \geq 7/4$~\cite{lubetzky2012critical}, which coincides with the bound \eqref{previous}.

In this work, we present a rigorous improvement of this bound for the kinetic Ising model in any dimension $d\geq2$:
\begin{align}
z \geq 2
\end{align}
using Eq.~\eqref{defz2} as the definition of $z$. This bound for the two-dimensional Ising model was conjectured before in Ref.~\cite{PhysRevB.83.125114}.
The key step is our recent work, which proved $z \geq 2$ quite generally for continuous-time Markov chains corresponding to a statistical mechanics model, assuming only the detailed balance condition and the locality of the update rule~\cite{masaokaMCMC}.  
This work presents a shorter and mathematically more careful version of the proof focusing on the Ising model.

Our proof is based on a lower-bound on correlation functions  that follows from the correlation inequality due to Simon and Lieb~\cite{Boel,Simon,Lieb} and an upper-bound on correlation function by Gosset and Huang~\cite{PhysRevLett.116.097202}. Our proof not only applies to the critical point, but also to any equilibrium state in which the two-point correlation function does not exhibit exponential decay.

\section{Definitions}
\subsection{Ising model in equilibrium}
In this work, we consider the standard ferromagnetic Ising model with the nearest neighbor interaction on the $d$-dimensional hypercubic lattice $\Lambda_L=\big\{-\frac{L-1}{2},-\frac{L-3}{2},\cdots,\frac{L-1}{2}\big\}^d\subset\mathbb{Z}^d$, 
where $L$ is odd. We denote the $a$-th component of $i$ by $i^{(a)}$ ($a=1,2,\cdots,d$).
For each site $i\in\Lambda_L$, we associate a spin variable $\sigma_i = \pm 1$. We denote by $\bm{\sigma}=(\sigma_i)_{i\in\Lambda_L}$ a
spin configuration, and by $\mathcal{S}_L$ the set of all spin configurations.

The energy of a spin configuration $\bm{\sigma}\in\mathcal{S}_L$ is given by
\begin{align}
E(\bm{\sigma})=-\sum_{(i,j)}\sigma_i\sigma_j=-\frac{1}{2}\sum_{i\in\Lambda_L}\sigma_i\sum_{j\in B_i}\sigma_j,
\end{align}
where the summation in the first expression is over nearest neighbor pairs, and $B_i$ in the second expression is the set of sites adjacent to the site $i$. We use periodic boundary conditions to regard $\Lambda_L$ as a torus.
For any function $O$ of spin configurations, we denote its expectation value in the equilibrium state at inverse temperature $\beta>0$ as
\begin{align}
\langle O\rangle\coloneqq\frac{1}{Z}\sum_{\bm{\sigma}\in\mathcal{S}_L}O(\bm{\sigma})w(\bm{\sigma}),\label{exp}
\end{align}
where
\begin{align}
w(\bm{\sigma})\coloneqq e^{-\beta E(\bm{\sigma})}>0\label{isingbw}
\end{align}
is the Boltzmann weight and
\begin{align}
Z\coloneqq\sum_{\bm{\sigma}\in\mathcal{S}_L}w(\bm{\sigma})
\end{align}
is the partition function. We note that the spin flipping symmetry $w(\bm{\sigma})=w(-\bm{\sigma})$ implies $\langle \sigma_i\rangle=0$. Also, the first Griffiths inequality~\cite{10.1063/1.1705219} states
\begin{align}
0\leq\langle \sigma_i\sigma_j\rangle\leq1.\label{correlationineq}
\end{align}
It is well known that the model undergoes a ferromagnetic phase transition if $d\geq2$~\cite
{FriedliVelenik,SimonBook}. It has been proved that 
\begin{align}
\limsup_{L\to\infty}\sum_{i\in\Lambda_{L}}\langle\sigma_o\sigma_i\rangle_{L}
\begin{cases}
<\infty&\text{if $\beta<\beta_c$};\\
=\infty&\text{if $\beta\geq\beta_c$},
\end{cases}
\end{align}
where $o\in\Lambda_L$ is the origin $(0,0,\cdots,0)$ and $\beta_c<\infty$ is the critical inverse temperature.

\subsection{Kinetic Ising model}
Now we introduce the kinetic Ising model, also called the stochastic Ising model, which implements the Glauber dynamics.
Let $p(t,\bm{\sigma})\geq0$ be the probability for the spin configuration $\bm{\sigma}$ to be realized at time $t$. We consider a continuous-time Markov process described by the master equation\footnote{See, for example, Ref.~\cite{shiraishi} for a general introduction to stochastic processes.}
\begin{align}
\frac{d}{dt}p(t,\bm{\sigma})=\sum_{\bm{\sigma}'\in\mathcal{S}_L}W_{\bm{\sigma},\bm{\sigma}'}p(t,\bm{\sigma}').
\end{align}
Let $P_i(\bm{\sigma})\geq0$ be the rate at
which the spin at $i$ is flipped. Let us denote by $\tau_i(\bm{\sigma})$ the spin configuration obtained by flipping the spin at site $i$ in $\bm{\sigma}\in \mathcal{S}_L$. Then the transition rate matrix $W$ is given by $\sum_{i\in\Lambda_L}W_i$, where
\begin{align}
(W_i)_{\bm{\sigma},\bm{\sigma}'}=
\begin{cases}
P_i(\bm{\sigma}')&(\bm{\sigma}=\tau_i(\bm{\sigma}'))\\
-P_i(\bm{\sigma}')&(\bm{\sigma}=\bm{\sigma}')\\
0&(\text{otherwise})
\end{cases},\label{defWP}
\end{align}
which satisfies
\begin{align}
\sum_{\bm{\sigma}\in\mathcal{S}_L}(W_i)_{\bm{\sigma},\bm{\sigma}'}=0.\label{conservation}
\end{align}
This relation is necessary for the conservation of probability $\sum_{\bm{\sigma}\in\mathcal{S}_L}p(t,\bm{\sigma})=1$.

For concreteness, in this work we assume the heat-bath algorithm~\cite{10.1143/PTP.39.947}
\begin{align}
P_i(\bm{\sigma})\coloneqq\frac{1}{2}-\frac{1}{2}\sigma_i\tanh\Big(\beta\sum_{j\in B_i}\sigma_j\Big)>0,\label{heatbath}
\end{align}
which satisfies the detailed balance condition
\begin{align}
(W_i)_{\bm{\sigma},\bm{\sigma}'}w(\bm{\sigma}')=(W_i)_{\bm{\sigma}',\bm{\sigma}}w(\bm{\sigma})\label{detailed}
\end{align}
for any $\bm{\sigma},\bm{\sigma}'\in\mathcal{S}_L$. Other choices of $P_i(\bm{\sigma})$ can be treated in the same way with minor modifications as discussed in Appendix~\ref{Generalization}. The Boltzmann weight $w(\bm{\sigma})>0$ is an eigenvector of $W$ with the eigenvalue $0$:
\begin{align}
\sum_{\bm{\sigma}'\in\mathcal{S}_L}(W_i)_{\bm{\sigma},\bm{\sigma}'}w(\bm{\sigma}')=\Big(\sum_{\bm{\sigma}'\in\mathcal{S}_L}(W_i)_{\bm{\sigma}',\bm{\sigma}}\Big)w(\bm{\sigma})=0,\label{BW}
\end{align}
As we will see in Sec.~\ref{secRSM}, $w(\bm{\sigma})/Z$ is the unique stationary distribution of our Markov process.

\subsection{Real symmetric matrices}
\label{secRSM}
In order to relate the Markov process to a quantum mechanical problem, we use the standard similarity transformation to define a matrix $H_i$ for each $i \in\Lambda_L$ by
\begin{align}
(H_i)_{\bm{\sigma},\bm{\sigma}'}\coloneqq-\sqrt{\frac{w(\bm{\sigma}')}{w(\bm{\sigma})}}(W_i)_{\bm{\sigma},\bm{\sigma}'}.\label{HW}
\end{align}
The matrix will be used in \eqref{RKH} to define the (generalized) Rokhsar-Kivelson Hamiltonian~\cite{CLHenley_2004,CASTELNOVO2005316,PhysRevB.83.125114,PhysRevLett.96.220601}.
Note that the detailed balance condition \eqref{detailed} implies $(H_i)_{\bm{\sigma},\bm{\sigma}'}=(H_i)_{\bm{\sigma}',\bm{\sigma}}$ and hence $H_i$ is a real symmetric
matrix. Since $H_i$ and $-W_i$ are related to each other by the similarity transformation in \eqref{HW}, their eigenvalues are common. 
Also, the non-positive nature of off-diagonal components of $H_i$, combined with Eq.~\eqref{BW}, implies that $H_i$ is positive semi-definite~\footnote{Let $v$ be an eigenvector of $H_i$ with the lowest eigenvalue $\mu_i$ and $u$ be a vector whose components are given by $u_{\bm{\sigma}}=|v_{\bm{\sigma}}|$. Since off-diagonal components of $H_i$ are nonpositive, $\mu_i=\sum_{\bm{\sigma},\bm{\sigma}'}v_{\bm{\sigma}}(H_i)_{\bm{\sigma},\bm{\sigma}'}v_{\bm{\sigma}'}\geq\sum_{\bm{\sigma},\bm{\sigma}'}u_{\bm{\sigma}}(H_i)_{\bm{\sigma},\bm{\sigma}'}u_{\bm{\sigma}'}$. Then the variational principle implies that $u_{\bm{\sigma}}$ is also an eigenvector of $H_i$ with the eigenvalue $\mu_i$. On the other hand, $w_{\bm{\sigma}}=\sqrt{w(\bm{\sigma})}>0$ is an eigenvector of $H_i$ with the eigenvalue $0$ and $\sum_{\bm{\sigma}\in\mathcal{S}_L}w_{\bm{\sigma}}u_{\bm{\sigma}}=\sum_{\bm{\sigma}\in\mathcal{S}_L}w_{\bm{\sigma}}|v_{\bm{\sigma}}|\neq0$. Therefore, $\mu_i=0$.}. 
We also find from \eqref{BW} that the normalized vector $\bm{\Phi}_0$ defined by
\begin{align}
(\bm{\Phi}_0)_{\bm{\sigma}}\coloneqq\sqrt{\frac{w(\bm{\sigma})}{Z}}
\end{align}
is an eigenvector of $H_i$ with eigenvalue $0$. It should be noted that the vector $\bm{\Phi}_0$ does not depend on $i$. As long as $0\leq \beta<\infty$, the Perron--Frobenius theorem~\cite{tasaki2020physics} implies that the eigenvalue 0 of $H=\sum_{i\in\Lambda_L}H_i$ is nondegenerate. Therefore, we have $\lim_{t\to\infty}(e^{-Ht})_{\bm{\sigma},\bm{\sigma}'}=\sqrt{w(\bm{\sigma})w(\bm{\sigma}')}/Z$.

Let us denote by $\epsilon_L$ the smallest eigenvalue of $-W$ other than $0$. The quantity $\epsilon_L$, which is called the spectral gap in the mathematical literature, essentially determines the decay property of
autocorrelation functions. For example, the equilibrium autocorrelation function of an operator $O$ is defined by
\begin{align}
\langle Oe^{Wt}O\rangle=\frac{1}{Z}\sum_{\bm{\sigma},\bm{\sigma}'\in\mathcal{S}_L}O(\bm{\sigma})(e^{Wt})_{\bm{\sigma},\bm{\sigma}'}O(\bm{\sigma}')w(\bm{\sigma}').
\end{align}
For any $\bm{\sigma}'$, $\lim_{t\to\infty}(e^{Wt})_{\bm{\sigma},\bm{\sigma}'}=w(\bm{\sigma})/Z$, which implies
\begin{align}
\lim_{t\to\infty}\langle Oe^{Wt}O\rangle&=\sum_{\bm{\sigma}\in\mathcal{S}_L}O(\bm{\sigma})\frac{w(\bm{\sigma})}{Z}\sum_{\bm{\sigma}'\in\mathcal{S}_L}O(\bm{\sigma}')\frac{w(\bm{\sigma}')}{Z}\notag\\
&=\langle O\rangle^2.
\end{align}
Then the autocorrelation function behaves as
\begin{align}
\big|\langle O e^{Wt}O\rangle-\langle O\rangle^2\big|\simeq C e^{-\epsilon_L t}.\label{OeWO}
\end{align}

Suppose that the relaxation time $\tau=1/\epsilon_L$ grows with $L$ as $L^z$. This scaling defines the dynamical critical exponent $z$.

\subsection{Theorem}
Our main theorem is the following upper bound for the spectral gap $\epsilon_L$.

\textbf{Theorem}
\textit{Let $d\geq2$. Then there exist constants $C$ and $L_0$ that depend only on $d$ such that 
\begin{align}
0<\epsilon_L\leq C\Big(\frac{\log L}{L}\Big)^2\label{statement}
\end{align}
for any $\beta\geq\beta_c$ and $L\geq L_0$.}

Comparing the bound \eqref{statement} at $\beta=\beta_c$ with the expected finite size scaling behavior $\epsilon_L \simeq L^{-z}$, we see that the theorem implies $z\geq2$ (assuming the existence of the exponent $z$). We also note that the bound \eqref{statement} is most meaningful for $\beta=\beta_c$ since it is expected that $\epsilon_L$ decays much faster as $L$ grows in the low-temperature phase with $\beta>\beta_c$. See Ref.~\cite{Thomas} for known rigorous results for sufficiently large $\beta$.

\section{Proof}
We examine the behavior of the quantity
\begin{align}
c(L)\coloneqq 2d\sum_{j\in\partial\Lambda_{L}}\langle \sigma_o\sigma_j\rangle\label{deflambda}
\end{align}
as a function of $L$, where $\partial\Lambda_{L}$ is the set of lattice sites in which at least one of its components satisfies $|i^{(a)}|=\frac{L-1}{2}$.  Note that $\partial\Lambda_{L}$ does not represent the boundary since we are imposing periodic boundary conditions.

We shall prove a lower bound \eqref{lower} and an upper bound \eqref{upper} for the same quantity $c(L)$.  The combination of the two bounds leads to the desired upper bound for the spectral gap.

\subsection{Lower bound}
A lower bound of $c(L)$ can be derived based on the Simon-Lieb inequality~\cite{Boel,Simon,Lieb}. If the correlation function $\langle \sigma_i\sigma_j\rangle_{L'}$ does not decay exponentially, then
\begin{align}
c(L)\geq1\label{lower}
\end{align}
for any $L=1,3,5,\cdots$~\cite{Boel,Simon,Lieb} . For readers' convenience, we review the proof in the Appendix~\ref{app:LB}.

\subsection{Upper bound}
In order to make use of results from quantum information, we shall identify the matrices discussed in the previous section with quantum mechanical operators. For each $i\in\Lambda_L$, let the local Hilbert space $\mathcal{H}_i$ be the two-dimensional Hilbert space with the basis $\{|+1\rangle_i, |-1\rangle_i\}$.
We consider a quantum system with the Hilbert space $\otimes_{i\in\Lambda_L}\mathcal{H}_i$ and denote its basis states as
\begin{align}
|\bm{\sigma}\rangle=\otimes_{i\in\Lambda_L}|\sigma_i\rangle_i,
\end{align}
where $\bm{\sigma}\in\mathcal{S}_L$.
Given $(H_i)_{\bm{\sigma},\bm{\sigma}'}$ in \eqref{HW}, we define the Rokhsar-Kivelson Hamiltonian $\hat{H}=\sum_{i\in\Lambda_{L}}\hat{H}_i$ by
\begin{align}
&\hat{H}_i\coloneqq\sum_{\bm{\sigma},\bm{\sigma}'\in\mathcal{S}_L}(H_i)_{\bm{\sigma},\bm{\sigma}'}|\bm{\sigma}\rangle\langle\bm{\sigma}'|.\label{RKH}
\end{align}
As detailed in the previous section, the spectrum of this Hamiltonian coincides with that of the transition rate matrix.

For the heat-bath algorithm \eqref{heatbath}, we find
\begin{align}
\hat{H}_i&=\frac{1}{2\cosh(\beta\sum_{j\in B_i}\hat{\sigma}_j^z)}\Big(e^{-\beta\hat{\sigma}_i^z{}\sum_{j\in B_i}\hat{\sigma}_j^z}-\hat{\sigma}_i^x\Big),\label{heatbath2}
\end{align}
which includes up to $(2d+1)$-spin interactions. Here $\hat{\sigma}_i^x$ and $\hat{\sigma}_i^z$ are operators defined by
\begin{align}
\hat{\sigma}_i^x=\sum_{\bm{\sigma}\in\mathcal{S}_L}|\tau_i(\bm{\sigma})\rangle\langle\bm{\sigma}|,\quad\hat{\sigma}_i^z=\sum_{\bm{\sigma}\in\mathcal{S}_L}\sigma_i|\bm{\sigma}\rangle\langle\bm{\sigma}|,
\end{align}
which can be represented by Pauli matrices
\begin{align}
\sigma^x=
\begin{pmatrix}
0&1\\
1&0
\end{pmatrix},\quad
\sigma^z=
\begin{pmatrix}
1&0\\
0&-1
\end{pmatrix}
\end{align}
on $\mathcal{H}_i$.
It is easily checked that $[\hat{H}_i,\hat{H}_j]=0$ unless $|i-j|=1$. 
As one can easily show, $\hat{H}_i^2=\hat{H}_i$, implying that $\hat{H}_i$ is a projector. Furthermore, 
\begin{align}
|\Phi_0\rangle\coloneqq\sum_{\bm{\sigma}\in\mathcal{S}_L}(\bm{\Phi}_0)_{\bm{\sigma}}|\bm{\sigma}\rangle\label{GS}
\end{align}
is a simultaneous ground state of all $\hat{H}_i$'s with the eigenvalue $0$:
\begin{align}
\hat{H}_i|\Phi_0\rangle=-\sum_{\bm{\sigma}\in\mathcal{S}_L}|\bm{\sigma}\rangle(H_i\bm{\Phi}_0)_{\bm{\sigma}}=0.\label{HiPhi}
\end{align}
In general, $\hat{H}=\sum_i\hat{H}_i$ is said to be frustration-free if a ground state $|\Phi\rangle$ of $\hat{H}$ is a simultaneous ground state of all $\hat{H}_i$'s. 
\eqref{HiPhi} implies that $\hat{H}$ is frustration-free.  

For a general frustration-free Hamiltonian $\hat{H}=\sum_{i}\hat{H}_i$ with $\hat{H}_i^2=\hat{H}_i$ and for arbitrary operators $\hat{O}$ and $\hat{O}'$, the following inequality holds~\cite{PhysRevLett.116.097202}\footnote{This expression improves the original inequality in Ref.~\cite{PhysRevLett.116.097202} in several ways: (i) the operator norm $\|\hat{O}\|$ is replaced by $\|\hat{O}|\Phi_0\rangle\|\leq \|\hat{O}\|$, (ii) $2c-1$ in the exponent is replaced by $2(c-1)$, and (iii) the definition of the distance between $\hat{O}$ and $\hat{O}'$ is refined. We include the proof of this version in the Appendix~\ref{app:GH}}:
\begin{align}
\label{GH}
&\frac{\langle\Phi_0|\hat{O}(1-\hat G)\hat{O}'|\Phi_0\rangle}{\|\hat{O}^\dagger|\Phi_0\rangle\| \|\hat{O}'|\Phi_0\rangle\|}\notag\\
&\leq 2e^2 \exp\Big(-\frac{D(\hat{O},\hat{O}')-1}{c-1}\sqrt{\frac{\epsilon_L}{g^2+\epsilon_L}}\Big), 
\end{align}
where $\hat{G}$ is the projector onto the ground states, $\epsilon_L$ is the smallest eigenvalue of $\hat{H}$ apart from $0$, and $D(\hat{O},\hat{O}')$ is a ``distance" between $\hat{O}$ and $\hat{O}'$ with respect to the Hamiltonian $\hat{H}$. See Appendix \ref{app:GH} for the definition.  For our Rokhsar-Kivelson Hamiltonian, $D(\hat{\sigma}_o^z,\hat{\sigma}_i^z)$ coincides with the $L^1$-norm $\|i\|_1$ under periodic boundary condition defined by $\|i\|_1\coloneqq\sum_{a=1}^d\min\big\{|i^{(a)}|,L-|i^{(a)}|\big\}$. We also have $c=2$, $g=2d$, $\hat{G}=|\Phi_0\rangle\langle\Phi_0|$, $\langle\Phi_0|\hat{\sigma}_o^z|\Phi_0\rangle=\langle\Phi_0|\hat{\sigma}_i^z|\Phi_0\rangle=0$, and $\|\hat{\sigma}_o^z|\Phi_0\rangle\|=\|\hat{\sigma}_i^z|\Phi_0\rangle\|\leq1$.
Plugging these in, we find\footnote{The inequality \eqref{GH} for $L\to\infty$ in the disordered phase directly implies that the correlation length $\xi$ is smaller than or equal to $\sqrt{\frac{(2d)^2+\epsilon_L}{\epsilon_L}}$. This suggests that $\tau=\frac{1}{\epsilon_L}\geq \frac{\xi^2-1}{(2d)^2}$. Hence, if one uses Eq.~\eqref{defz1} as the definition of $z$, then this relation already implies that $z\geq2$ for the critical point.}
\begin{align}
\langle\Phi_0|\hat{\sigma}_o^z\hat{\sigma}_i^z|\Phi_0\rangle\leq 2e^2 \exp\Big(-(\|i\|_1-1)\sqrt{\frac{\epsilon_L}{(2d)^2+\epsilon_L}}\Big).\label{GH}
\end{align}
Since $\langle\Phi_0|\hat{\sigma}_o^z\hat{\sigma}_i^z|\Phi_0\rangle=\frac{1}{Z}\sum_{\bm{\sigma}\in\mathcal{S}_L}\sigma_o\sigma_iw(\bm{\sigma})=\langle\sigma_o\sigma_i\rangle$, the inequality \eqref{GH} gives an upper bound of $c(L)$:
\begin{align}
c(L)\leq 2d|\partial\Lambda_{L}|\cdot 2e^2\exp\Big(-\frac{L-3}{2}\sqrt{\frac{\epsilon_L}{(2d)^2+\epsilon_L}}\Big),\label{upper}
\end{align}
where $|\partial\Lambda_{L}|\leq 2dL^{d-1}$.

\subsection{Proof of \eqref{statement}}

Inequalities \eqref{lower} and  \eqref{upper} imply
\begin{align}
\frac{\epsilon_L}{\epsilon_L+(2d)^2} \leq f(L)^2,\label{statement1}
\end{align}
where
\begin{align}
f(L)\coloneqq \frac{2}{L-3} \ln(8e^2d^2L^{d-1}).
\end{align}
When $f(L)^2<1$, this can be written as
\begin{align}
0<\epsilon_L \leq (2d)^2\frac{f(L)^2}{1-f(L)^2}\label{statement2}.
\end{align}
We find that
\begin{align}
0<\epsilon_L\leq (2d)^4\Big(\frac{\log L}{L}\Big)^2
\end{align}
for any $L\geq 10e^2d^2$.

\section{Discussions}
Our results are applicable not only to the critical point but also to the entire ordered phase.
In the ordered phase, $\epsilon_L$ measures the finite-size gap of quasi-degenerate ground states, which is typically exponentially small.
A conjecture in Ref.~\cite{PhysRevB.110.195140} suggests that excitations are gapless throughout the ordered phase.
Furthermore, our methods naturally extend to broader classes of Markov chains; see Ref.~\cite{masaokaMCMC} for a comprehensive discussion.

Before closing, we note that Refs.~\cite{gossetLocalGapThreshold2016,anshuImprovedLocalSpectral2020,lemmQuantitativelyImprovedFinitesize2022,lemmCriticalFinitesizeGap2024} establish upper bounds on the spectral gap for frustration-free Hamiltonians under \emph{open} boundary conditions without boundary terms.\footnote{Under the open boundary condition without boundary terms, the Hamiltonian decomposes into a direct sum of sectors, each defined by fixing the boundary spins in every possible configuration. This situation differs from the familiar plus/minus boundaries, where all boundary spins are fixed to $+1$ or $-1$, and from free boundaries, where the boundary spins remain completely unconstrained.}
The bound $\epsilon_L \le C/L^2$ derived in the first three works is not directly applicable to our model~\eqref{heatbath2}, as it assumes strictly nearest-neighbor interactions. 
The result in Ref.~\cite{lemmCriticalFinitesizeGap2024} implies a similar bound, but there are important differences: it considers anisotropic shapes in addition to open boundary conditions and focuses on the asymptotic behavior in the large $L$ limit, rather than providing a bound for each finite system. 
Hence, the conclusion is not directly comparable to ours.

\begin{acknowledgments}
We thank Hal Tasaki for valuable comments on the earlier version of the manuscript. 
We thank Carolyn Zhang and Ruben Verresen for useful discussions.
The work of H.W. is supported by JSPS KAKENHI Grant No. JP24K00541.
This research is funded in part by the
Gordon and Betty Moore Foundation's EPiQS Initiative,
Grant GBMF8683 to T.S.
\end{acknowledgments}

\bibliography{ref}

\begin{thebibliography}{40}%
\makeatletter
\providecommand \@ifxundefined [1]{%
 \@ifx{#1\undefined}
}%
\providecommand \@ifnum [1]{%
 \ifnum #1\expandafter \@firstoftwo
 \else \expandafter \@secondoftwo
 \fi
}%
\providecommand \@ifx [1]{%
 \ifx #1\expandafter \@firstoftwo
 \else \expandafter \@secondoftwo
 \fi
}%
\providecommand \natexlab [1]{#1}%
\providecommand \enquote  [1]{``#1''}%
\providecommand \bibnamefont  [1]{#1}%
\providecommand \bibfnamefont [1]{#1}%
\providecommand \citenamefont [1]{#1}%
\providecommand \href@noop [0]{\@secondoftwo}%
\providecommand \href [0]{\begingroup \@sanitize@url \@href}%
\providecommand \@href[1]{\@@startlink{#1}\@@href}%
\providecommand \@@href[1]{\endgroup#1\@@endlink}%
\providecommand \@sanitize@url [0]{\catcode `\\12\catcode `\$12\catcode
  `\&12\catcode `\#12\catcode `\^12\catcode `\_12\catcode `\%12\relax}%
\providecommand \@@startlink[1]{}%
\providecommand \@@endlink[0]{}%
\providecommand \url  [0]{\begingroup\@sanitize@url \@url }%
\providecommand \@url [1]{\endgroup\@href {#1}{\urlprefix }}%
\providecommand \urlprefix  [0]{URL }%
\providecommand \Eprint [0]{\href }%
\providecommand \doibase [0]{http://dx.doi.org/}%
\providecommand \selectlanguage [0]{\@gobble}%
\providecommand \bibinfo  [0]{\@secondoftwo}%
\providecommand \bibfield  [0]{\@secondoftwo}%
\providecommand \translation [1]{[#1]}%
\providecommand \BibitemOpen [0]{}%
\providecommand \bibitemStop [0]{}%
\providecommand \bibitemNoStop [0]{.\EOS\space}%
\providecommand \EOS [0]{\spacefactor3000\relax}%
\providecommand \BibitemShut  [1]{\csname bibitem#1\endcsname}%
\let\auto@bib@innerbib\@empty
%</preamble>
\bibitem [{\citenamefont {Hasenbusch}\ \emph {et~al.}(1999)\citenamefont
  {Hasenbusch}, \citenamefont {Pinn},\ and\ \citenamefont
  {Vinti}}]{hasenbuschCriticalExponentsThreedimensional1999}%
  \BibitemOpen
  \bibfield  {author} {\bibinfo {author} {\bibfnamefont {M.}~\bibnamefont
  {Hasenbusch}}, \bibinfo {author} {\bibfnamefont {K.}~\bibnamefont {Pinn}}, \
  and\ \bibinfo {author} {\bibfnamefont {S.}~\bibnamefont {Vinti}},\ }\href
  {\doibase 10.1103/PhysRevB.59.11471} {\bibfield  {journal} {\bibinfo
  {journal} {Physical Review B}\ }\textbf {\bibinfo {volume} {59}},\ \bibinfo
  {pages} {11471} (\bibinfo {year} {1999})}\BibitemShut {NoStop}%
\bibitem [{\citenamefont {Poland}\ and\ \citenamefont
  {Simmons-Duffin}(2016)}]{Poland}%
  \BibitemOpen
  \bibfield  {author} {\bibinfo {author} {\bibfnamefont {D.}~\bibnamefont
  {Poland}}\ and\ \bibinfo {author} {\bibfnamefont {D.}~\bibnamefont
  {Simmons-Duffin}},\ }\href {\doibase 10.1038/nphys3761} {\bibfield  {journal}
  {\bibinfo  {journal} {Nature Physics}\ }\textbf {\bibinfo {volume} {12}},\
  \bibinfo {pages} {535} (\bibinfo {year} {2016})}\BibitemShut {NoStop}%
\bibitem [{\citenamefont
  {Onsager}(1944)}]{onsagerCrystalStatisticsTwoDimensional1944}%
  \BibitemOpen
  \bibfield  {author} {\bibinfo {author} {\bibfnamefont {L.}~\bibnamefont
  {Onsager}},\ }\href {\doibase 10.1103/PhysRev.65.117} {\bibfield  {journal}
  {\bibinfo  {journal} {Physical Review}\ }\textbf {\bibinfo {volume} {65}},\
  \bibinfo {pages} {117} (\bibinfo {year} {1944})}\BibitemShut {NoStop}%
\bibitem [{\citenamefont {Fern{\'a}ndez}\ \emph {et~al.}(1992)\citenamefont
  {Fern{\'a}ndez}, \citenamefont {Fr{\"o}hlich},\ and\ \citenamefont
  {Sokal}}]{fernandez}%
  \BibitemOpen
  \bibfield  {author} {\bibinfo {author} {\bibfnamefont {R.}~\bibnamefont
  {Fern{\'a}ndez}}, \bibinfo {author} {\bibfnamefont {J.}~\bibnamefont
  {Fr{\"o}hlich}}, \ and\ \bibinfo {author} {\bibfnamefont {A.~D.}\
  \bibnamefont {Sokal}},\ }\href@noop {} {\emph {\bibinfo {title} {Random
  Walks, Critical Phenomena, and Triviality in Quantum Field Theory}}}\
  (\bibinfo  {publisher} {Springer},\ \bibinfo {address} {Berlin},\ \bibinfo
  {year} {1992})\BibitemShut {NoStop}%
\bibitem [{\citenamefont {Halperin}(1973)}]{PhysRevB.8.4437}%
  \BibitemOpen
  \bibfield  {author} {\bibinfo {author} {\bibfnamefont {B.~I.}\ \bibnamefont
  {Halperin}},\ }\href {\doibase 10.1103/PhysRevB.8.4437} {\bibfield  {journal}
  {\bibinfo  {journal} {Phys. Rev. B}\ }\textbf {\bibinfo {volume} {8}},\
  \bibinfo {pages} {4437} (\bibinfo {year} {1973})}\BibitemShut {NoStop}%
\bibitem [{\citenamefont {Abe}\ and\ \citenamefont
  {Hatano}(1969)}]{10.1143/PTP.41.941}%
  \BibitemOpen
  \bibfield  {author} {\bibinfo {author} {\bibfnamefont {R.}~\bibnamefont
  {Abe}}\ and\ \bibinfo {author} {\bibfnamefont {A.}~\bibnamefont {Hatano}},\
  }\href {\doibase 10.1143/PTP.41.941} {\bibfield  {journal} {\bibinfo
  {journal} {Prog. Theor. Phys.}\ }\textbf {\bibinfo {volume} {41}},\ \bibinfo
  {pages} {941} (\bibinfo {year} {1969})}\BibitemShut {NoStop}%
\bibitem [{Note1()}]{Note1}%
  \BibitemOpen
  \bibinfo {note} {We review the discussion in Ref.~\cite {PhysRevB.8.4437} in
  Appendix~\ref {sec:Halperin}.}\BibitemShut {Stop}%
\bibitem [{\citenamefont {Nightingale}\ and\ \citenamefont
  {Bl\"ote}(2000)}]{PhysRevB.62.1089}%
  \BibitemOpen
  \bibfield  {author} {\bibinfo {author} {\bibfnamefont {M.~P.}\ \bibnamefont
  {Nightingale}}\ and\ \bibinfo {author} {\bibfnamefont {H.~W.~J.}\
  \bibnamefont {Bl\"ote}},\ }\href {\doibase 10.1103/PhysRevB.62.1089}
  {\bibfield  {journal} {\bibinfo  {journal} {Phys. Rev. B}\ }\textbf {\bibinfo
  {volume} {62}},\ \bibinfo {pages} {1089} (\bibinfo {year}
  {2000})}\BibitemShut {NoStop}%
\bibitem [{\citenamefont {Hasenbusch}(2020)}]{PhysRevE.101.022126}%
  \BibitemOpen
  \bibfield  {author} {\bibinfo {author} {\bibfnamefont {M.}~\bibnamefont
  {Hasenbusch}},\ }\href {\doibase 10.1103/PhysRevE.101.022126} {\bibfield
  {journal} {\bibinfo  {journal} {Phys. Rev. E}\ }\textbf {\bibinfo {volume}
  {101}},\ \bibinfo {pages} {022126} (\bibinfo {year} {2020})}\BibitemShut
  {NoStop}%
\bibitem [{\citenamefont {Adzhemyan}\ \emph {et~al.}(2022)\citenamefont
  {Adzhemyan}, \citenamefont {Evdokimov}, \citenamefont {Hnati{\v c}},
  \citenamefont {Ivanova}, \citenamefont {Kompaniets}, \citenamefont {Kudlis},\
  and\ \citenamefont {Zakharov}}]{adzhemyanDynamicCriticalExponent2022}%
  \BibitemOpen
  \bibfield  {author} {\bibinfo {author} {\bibfnamefont {L.~{\relax Ts}.}\
  \bibnamefont {Adzhemyan}}, \bibinfo {author} {\bibfnamefont {D.~A.}\
  \bibnamefont {Evdokimov}}, \bibinfo {author} {\bibfnamefont {M.}~\bibnamefont
  {Hnati{\v c}}}, \bibinfo {author} {\bibfnamefont {E.~V.}\ \bibnamefont
  {Ivanova}}, \bibinfo {author} {\bibfnamefont {M.~V.}\ \bibnamefont
  {Kompaniets}}, \bibinfo {author} {\bibfnamefont {A.}~\bibnamefont {Kudlis}},
  \ and\ \bibinfo {author} {\bibfnamefont {D.~V.}\ \bibnamefont {Zakharov}},\
  }\href {\doibase 10.1016/j.physleta.2021.127870} {\bibfield  {journal}
  {\bibinfo  {journal} {Physics Letters A}\ }\textbf {\bibinfo {volume}
  {425}},\ \bibinfo {pages} {127870} (\bibinfo {year} {2022})}\BibitemShut
  {NoStop}%
\bibitem [{\citenamefont {Lubetzky}\ and\ \citenamefont
  {Sly}(2012)}]{lubetzky2012critical}%
  \BibitemOpen
  \bibfield  {author} {\bibinfo {author} {\bibfnamefont {E.}~\bibnamefont
  {Lubetzky}}\ and\ \bibinfo {author} {\bibfnamefont {A.}~\bibnamefont {Sly}},\
  }\href {\doibase 10.1007/s00220-012-1460-9} {\bibfield  {journal} {\bibinfo
  {journal} {Commun. Math. Phys.}\ }\textbf {\bibinfo {volume} {313}},\
  \bibinfo {pages} {815} (\bibinfo {year} {2012})}\BibitemShut {NoStop}%
\bibitem [{\citenamefont {Isakov}\ \emph {et~al.}(2011)\citenamefont {Isakov},
  \citenamefont {Fendley}, \citenamefont {Ludwig}, \citenamefont {Trebst},\
  and\ \citenamefont {Troyer}}]{PhysRevB.83.125114}%
  \BibitemOpen
  \bibfield  {author} {\bibinfo {author} {\bibfnamefont {S.~V.}\ \bibnamefont
  {Isakov}}, \bibinfo {author} {\bibfnamefont {P.}~\bibnamefont {Fendley}},
  \bibinfo {author} {\bibfnamefont {A.~W.~W.}\ \bibnamefont {Ludwig}}, \bibinfo
  {author} {\bibfnamefont {S.}~\bibnamefont {Trebst}}, \ and\ \bibinfo {author}
  {\bibfnamefont {M.}~\bibnamefont {Troyer}},\ }\href {\doibase
  10.1103/PhysRevB.83.125114} {\bibfield  {journal} {\bibinfo  {journal} {Phys.
  Rev. B}\ }\textbf {\bibinfo {volume} {83}},\ \bibinfo {pages} {125114}
  (\bibinfo {year} {2011})}\BibitemShut {NoStop}%
\bibitem [{\citenamefont {Masaoka}\ \emph
  {et~al.}(2024{\natexlab{a}})\citenamefont {Masaoka}, \citenamefont
  {Soejima},\ and\ \citenamefont {Watanabe}}]{masaokaMCMC}%
  \BibitemOpen
  \bibfield  {author} {\bibinfo {author} {\bibfnamefont {R.}~\bibnamefont
  {Masaoka}}, \bibinfo {author} {\bibfnamefont {T.}~\bibnamefont {Soejima}}, \
  and\ \bibinfo {author} {\bibfnamefont {H.}~\bibnamefont {Watanabe}},\ }\href
  {\doibase 10.48550/arXiv.2406.06415} {\  (\bibinfo {year}
  {2024}{\natexlab{a}}),\ 10.48550/arXiv.2406.06415}\BibitemShut {NoStop}%
\bibitem [{\citenamefont {Boel}\ and\ \citenamefont {Kasteleyn}(1978)}]{Boel}%
  \BibitemOpen
  \bibfield  {author} {\bibinfo {author} {\bibfnamefont {R.~J.}\ \bibnamefont
  {Boel}}\ and\ \bibinfo {author} {\bibfnamefont {P.~W.}\ \bibnamefont
  {Kasteleyn}},\ }\href {\doibase 10.1007/BF01940764} {\bibfield  {journal}
  {\bibinfo  {journal} {Commun. Math. Phys.}\ }\textbf {\bibinfo {volume}
  {61}},\ \bibinfo {pages} {191} (\bibinfo {year} {1978})}\BibitemShut
  {NoStop}%
\bibitem [{\citenamefont {Simon}(1980)}]{Simon}%
  \BibitemOpen
  \bibfield  {author} {\bibinfo {author} {\bibfnamefont {B.}~\bibnamefont
  {Simon}},\ }\href {\doibase 10.1007/BF01982711} {\bibfield  {journal}
  {\bibinfo  {journal} {Commun. Math. Phys.}\ }\textbf {\bibinfo {volume}
  {77}},\ \bibinfo {pages} {111} (\bibinfo {year} {1980})}\BibitemShut
  {NoStop}%
\bibitem [{\citenamefont {Lieb}(1980)}]{Lieb}%
  \BibitemOpen
  \bibfield  {author} {\bibinfo {author} {\bibfnamefont {E.~H.}\ \bibnamefont
  {Lieb}},\ }\href {\doibase 10.1007/BF01982712} {\bibfield  {journal}
  {\bibinfo  {journal} {Commun. Math. Phys.}\ }\textbf {\bibinfo {volume}
  {77}},\ \bibinfo {pages} {127} (\bibinfo {year} {1980})}\BibitemShut
  {NoStop}%
\bibitem [{\citenamefont {Gosset}\ and\ \citenamefont
  {Huang}(2016)}]{PhysRevLett.116.097202}%
  \BibitemOpen
  \bibfield  {author} {\bibinfo {author} {\bibfnamefont {D.}~\bibnamefont
  {Gosset}}\ and\ \bibinfo {author} {\bibfnamefont {Y.}~\bibnamefont {Huang}},\
  }\href {\doibase 10.1103/PhysRevLett.116.097202} {\bibfield  {journal}
  {\bibinfo  {journal} {Phys. Rev. Lett.}\ }\textbf {\bibinfo {volume} {116}},\
  \bibinfo {pages} {097202} (\bibinfo {year} {2016})}\BibitemShut {NoStop}%
\bibitem [{\citenamefont {Griffiths}(1967)}]{10.1063/1.1705219}%
  \BibitemOpen
  \bibfield  {author} {\bibinfo {author} {\bibfnamefont {R.~B.}\ \bibnamefont
  {Griffiths}},\ }\href {\doibase 10.1063/1.1705219} {\bibfield  {journal}
  {\bibinfo  {journal} {J. Math. Phys.}\ }\textbf {\bibinfo {volume} {8}},\
  \bibinfo {pages} {478} (\bibinfo {year} {1967})}\BibitemShut {NoStop}%
\bibitem [{\citenamefont {Friedli}\ and\ \citenamefont
  {Velenik}(2017)}]{FriedliVelenik}%
  \BibitemOpen
  \bibfield  {author} {\bibinfo {author} {\bibfnamefont {S.}~\bibnamefont
  {Friedli}}\ and\ \bibinfo {author} {\bibfnamefont {Y.}~\bibnamefont
  {Velenik}},\ }\href@noop {} {\emph {\bibinfo {title} {Statistical Mechanics
  of Lattice Systems: A Concrete Mathematical Introduction}}}\ (\bibinfo
  {publisher} {Cambridge University Press},\ \bibinfo {address} {Cambridge},\
  \bibinfo {year} {2017})\BibitemShut {NoStop}%
\bibitem [{\citenamefont {Simon}(1993)}]{SimonBook}%
  \BibitemOpen
  \bibfield  {author} {\bibinfo {author} {\bibfnamefont {B.}~\bibnamefont
  {Simon}},\ }\href@noop {} {\emph {\bibinfo {title} {The Statistical Mechanics
  of Lattice Gases, Volume I}}}\ (\bibinfo  {publisher} {Princeton University
  Press},\ \bibinfo {address} {Princeton},\ \bibinfo {year} {1993})\BibitemShut
  {NoStop}%
\bibitem [{Note2()}]{Note2}%
  \BibitemOpen
  \bibinfo {note} {See, for example, Ref.~\cite {shiraishi} for a general
  introduction to stochastic processes.}\BibitemShut {Stop}%
\bibitem [{\citenamefont {Abe}(1968)}]{10.1143/PTP.39.947}%
  \BibitemOpen
  \bibfield  {author} {\bibinfo {author} {\bibfnamefont {R.}~\bibnamefont
  {Abe}},\ }\href {\doibase 10.1143/PTP.39.947} {\bibfield  {journal} {\bibinfo
   {journal} {Prog. Theor. Phys.}\ }\textbf {\bibinfo {volume} {39}},\ \bibinfo
  {pages} {947} (\bibinfo {year} {1968})}\BibitemShut {NoStop}%
\bibitem [{\citenamefont {Henley}(2004)}]{CLHenley_2004}%
  \BibitemOpen
  \bibfield  {author} {\bibinfo {author} {\bibfnamefont {C.~L.}\ \bibnamefont
  {Henley}},\ }\href {\doibase 10.1088/0953-8984/16/11/045} {\bibfield
  {journal} {\bibinfo  {journal} {J. Phys.: Cond. Matt.}\ }\textbf {\bibinfo
  {volume} {16}},\ \bibinfo {pages} {S891} (\bibinfo {year}
  {2004})}\BibitemShut {NoStop}%
\bibitem [{\citenamefont {Castelnovo}\ \emph {et~al.}(2005)\citenamefont
  {Castelnovo}, \citenamefont {Chamon}, \citenamefont {Mudry},\ and\
  \citenamefont {Pujol}}]{CASTELNOVO2005316}%
  \BibitemOpen
  \bibfield  {author} {\bibinfo {author} {\bibfnamefont {C.}~\bibnamefont
  {Castelnovo}}, \bibinfo {author} {\bibfnamefont {C.}~\bibnamefont {Chamon}},
  \bibinfo {author} {\bibfnamefont {C.}~\bibnamefont {Mudry}}, \ and\ \bibinfo
  {author} {\bibfnamefont {P.}~\bibnamefont {Pujol}},\ }\href {\doibase
  https://doi.org/10.1016/j.aop.2005.01.006} {\bibfield  {journal} {\bibinfo
  {journal} {Annals of Physics}\ }\textbf {\bibinfo {volume} {318}},\ \bibinfo
  {pages} {316} (\bibinfo {year} {2005})}\BibitemShut {NoStop}%
\bibitem [{\citenamefont {Verstraete}\ \emph {et~al.}(2006)\citenamefont
  {Verstraete}, \citenamefont {Wolf}, \citenamefont {Perez-Garcia},\ and\
  \citenamefont {Cirac}}]{PhysRevLett.96.220601}%
  \BibitemOpen
  \bibfield  {author} {\bibinfo {author} {\bibfnamefont {F.}~\bibnamefont
  {Verstraete}}, \bibinfo {author} {\bibfnamefont {M.~M.}\ \bibnamefont
  {Wolf}}, \bibinfo {author} {\bibfnamefont {D.}~\bibnamefont {Perez-Garcia}},
  \ and\ \bibinfo {author} {\bibfnamefont {J.~I.}\ \bibnamefont {Cirac}},\
  }\href {\doibase 10.1103/PhysRevLett.96.220601} {\bibfield  {journal}
  {\bibinfo  {journal} {Phys. Rev. Lett.}\ }\textbf {\bibinfo {volume} {96}},\
  \bibinfo {pages} {220601} (\bibinfo {year} {2006})}\BibitemShut {NoStop}%
\bibitem [{Note3()}]{Note3}%
  \BibitemOpen
  \bibinfo {note} {Let $v$ be an eigenvector of $H_i$ with the lowest
  eigenvalue $\mu _i$ and $u$ be a vector whose components are given by
  $u_{\protect \bm {\sigma }}=|v_{\protect \bm {\sigma }}|$. Since off-diagonal
  components of $H_i$ are nonpositive, $\mu _i=\DOTSB \sum@ \slimits@
  _{\protect \bm {\sigma },\protect \bm {\sigma }'}v_{\protect \bm {\sigma
  }}(H_i)_{\protect \bm {\sigma },\protect \bm {\sigma }'}v_{\protect \bm
  {\sigma }'}\geq \DOTSB \sum@ \slimits@ _{\protect \bm {\sigma },\protect \bm
  {\sigma }'}u_{\protect \bm {\sigma }}(H_i)_{\protect \bm {\sigma },\protect
  \bm {\sigma }'}u_{\protect \bm {\sigma }'}$. Then the variational principle
  implies that $u_{\protect \bm {\sigma }}$ is also an eigenvector of $H_i$
  with the eigenvalue $\mu _i$. On the other hand, $w_{\protect \bm {\sigma
  }}=\protect \sqrt {w(\protect \bm {\sigma })}>0$ is an eigenvector of $H_i$
  with the eigenvalue $0$ and $\DOTSB \sum@ \slimits@ _{\protect \bm {\sigma
  }\in \protect \mathcal {S}_L}w_{\protect \bm {\sigma }}u_{\protect \bm
  {\sigma }}=\DOTSB \sum@ \slimits@ _{\protect \bm {\sigma }\in \protect
  \mathcal {S}_L}w_{\protect \bm {\sigma }}|v_{\protect \bm {\sigma }}|\protect
  \neq 0$. Therefore, $\mu _i=0$.}\BibitemShut {Stop}%
\bibitem [{\citenamefont {Tasaki}(2020)}]{tasaki2020physics}%
  \BibitemOpen
  \bibfield  {author} {\bibinfo {author} {\bibfnamefont {H.}~\bibnamefont
  {Tasaki}},\ }\href@noop {} {\emph {\bibinfo {title} {Physics and mathematics
  of quantum many-body systems}}},\ Vol.~\bibinfo {volume} {66}\ (\bibinfo
  {publisher} {Springer},\ \bibinfo {address} {Singapore},\ \bibinfo {year}
  {2020})\BibitemShut {NoStop}%
\bibitem [{\citenamefont {Thomas}(1989)}]{Thomas}%
  \BibitemOpen
  \bibfield  {author} {\bibinfo {author} {\bibfnamefont {L.~E.}\ \bibnamefont
  {Thomas}},\ }\href {\doibase 10.1007/BF02124328} {\bibfield  {journal}
  {\bibinfo  {journal} {Commun. Math. Phys.}\ }\textbf {\bibinfo {volume}
  {126}},\ \bibinfo {pages} {1} (\bibinfo {year} {1989})}\BibitemShut {NoStop}%
\bibitem [{Note4()}]{Note4}%
  \BibitemOpen
  \bibinfo {note} {This expression improves the original inequality in
  Ref.~\cite {PhysRevLett.116.097202} in several ways: (i) the operator norm
  $\|\protect \hat {O}\|$ is replaced by $\|\protect \hat {O}|\Phi _0\rangle
  \|\leq \|\protect \hat {O}\|$, (ii) $2c-1$ in the exponent is replaced by
  $2(c-1)$, and (iii) the definition of the distance between $\protect \hat
  {O}$ and $\protect \hat {O}'$ is refined. We include the proof of this
  version in the Appendix~\ref {app:GH}}\BibitemShut {NoStop}%
\bibitem [{Note5()}]{Note5}%
  \BibitemOpen
  \bibinfo {note} {The inequality \protect \eqref {GH} for $L\to \infty $ in
  the disordered phase directly implies that the correlation length $\xi $ is
  smaller than or equal to $\protect \sqrt {\protect \frac {(2d)^2+\epsilon
  _L}{\epsilon _L}}$. This suggests that $\tau =\protect \frac {1}{\epsilon
  _L}\geq \protect \frac {\xi ^2-1}{(2d)^2}$. Hence, if one uses Eq.~\protect
  \eqref {defz1} as the definition of $z$, then this relation already implies
  that $z\geq 2$ for the critical point.}\BibitemShut {Stop}%
\bibitem [{\citenamefont {Masaoka}\ \emph
  {et~al.}(2024{\natexlab{b}})\citenamefont {Masaoka}, \citenamefont
  {Soejima},\ and\ \citenamefont {Watanabe}}]{PhysRevB.110.195140}%
  \BibitemOpen
  \bibfield  {author} {\bibinfo {author} {\bibfnamefont {R.}~\bibnamefont
  {Masaoka}}, \bibinfo {author} {\bibfnamefont {T.}~\bibnamefont {Soejima}}, \
  and\ \bibinfo {author} {\bibfnamefont {H.}~\bibnamefont {Watanabe}},\ }\href
  {\doibase 10.1103/PhysRevB.110.195140} {\bibfield  {journal} {\bibinfo
  {journal} {Phys. Rev. B}\ }\textbf {\bibinfo {volume} {110}},\ \bibinfo
  {pages} {195140} (\bibinfo {year} {2024}{\natexlab{b}})}\BibitemShut
  {NoStop}%
\bibitem [{\citenamefont {Gosset}\ and\ \citenamefont
  {Mozgunov}(2016)}]{gossetLocalGapThreshold2016}%
  \BibitemOpen
  \bibfield  {author} {\bibinfo {author} {\bibfnamefont {D.}~\bibnamefont
  {Gosset}}\ and\ \bibinfo {author} {\bibfnamefont {E.}~\bibnamefont
  {Mozgunov}},\ }\href {\doibase 10.1063/1.4962337} {\bibfield  {journal}
  {\bibinfo  {journal} {J. Math. Phys.}\ }\textbf {\bibinfo {volume} {57}},\
  \bibinfo {pages} {091901} (\bibinfo {year} {2016})}\BibitemShut {NoStop}%
\bibitem [{\citenamefont {Anshu}(2020)}]{anshuImprovedLocalSpectral2020}%
  \BibitemOpen
  \bibfield  {author} {\bibinfo {author} {\bibfnamefont {A.}~\bibnamefont
  {Anshu}},\ }\href {\doibase 10.1103/PhysRevB.101.165104} {\bibfield
  {journal} {\bibinfo  {journal} {Physical Review B}\ }\textbf {\bibinfo
  {volume} {101}},\ \bibinfo {pages} {165104} (\bibinfo {year}
  {2020})}\BibitemShut {NoStop}%
\bibitem [{\citenamefont {Lemm}\ and\ \citenamefont
  {Xiang}(2022)}]{lemmQuantitativelyImprovedFinitesize2022}%
  \BibitemOpen
  \bibfield  {author} {\bibinfo {author} {\bibfnamefont {M.}~\bibnamefont
  {Lemm}}\ and\ \bibinfo {author} {\bibfnamefont {D.}~\bibnamefont {Xiang}},\
  }\href {\doibase 10.1088/1751-8121/ac7989} {\bibfield  {journal} {\bibinfo
  {journal} {J. Phys. A}\ }\textbf {\bibinfo {volume} {55}},\ \bibinfo {pages}
  {295203} (\bibinfo {year} {2022})}\BibitemShut {NoStop}%
\bibitem [{\citenamefont {Lemm}\ and\ \citenamefont
  {Lucia}(2024)}]{lemmCriticalFinitesizeGap2024}%
  \BibitemOpen
  \bibfield  {author} {\bibinfo {author} {\bibfnamefont {M.}~\bibnamefont
  {Lemm}}\ and\ \bibinfo {author} {\bibfnamefont {A.}~\bibnamefont {Lucia}},\
  }\href {\doibase 10.48550/arXiv.2409.09685} {\  (\bibinfo {year} {2024}),\
  10.48550/arXiv.2409.09685}\BibitemShut {NoStop}%
\bibitem [{Note6()}]{Note6}%
  \BibitemOpen
  \bibinfo {note} {Under the open boundary condition without boundary terms,
  the Hamiltonian decomposes into a direct sum of sectors, each defined by
  fixing the boundary spins in every possible configuration. This situation
  differs from the familiar plus/minus boundaries, where all boundary spins are
  fixed to $+1$ or $-1$, and from free boundaries, where the boundary spins
  remain completely unconstrained.}\BibitemShut {Stop}%
\bibitem [{\citenamefont {Shiraishi}(2023)}]{shiraishi}%
  \BibitemOpen
  \bibfield  {author} {\bibinfo {author} {\bibfnamefont {N.}~\bibnamefont
  {Shiraishi}},\ }\href@noop {} {\emph {\bibinfo {title} {An Introduction to
  Stochastic Thermodynamics}}}\ (\bibinfo  {publisher} {Springer},\ \bibinfo
  {address} {Singapore},\ \bibinfo {year} {2023})\BibitemShut {NoStop}%
\bibitem [{\citenamefont {Aharonov}\ \emph {et~al.}(2011)\citenamefont
  {Aharonov}, \citenamefont {Arad}, \citenamefont {Vazirani},\ and\
  \citenamefont {Landau}}]{aharonovDetectabilityLemmaIts2011}%
  \BibitemOpen
  \bibfield  {author} {\bibinfo {author} {\bibfnamefont {D.}~\bibnamefont
  {Aharonov}}, \bibinfo {author} {\bibfnamefont {I.}~\bibnamefont {Arad}},
  \bibinfo {author} {\bibfnamefont {U.}~\bibnamefont {Vazirani}}, \ and\
  \bibinfo {author} {\bibfnamefont {Z.}~\bibnamefont {Landau}},\ }\href
  {\doibase 10.1088/1367-2630/13/11/113043} {\bibfield  {journal} {\bibinfo
  {journal} {New Journal of Physics}\ }\textbf {\bibinfo {volume} {13}},\
  \bibinfo {pages} {113043} (\bibinfo {year} {2011})}\BibitemShut {NoStop}%
\bibitem [{\citenamefont {Anshu}\ \emph {et~al.}(2016)\citenamefont {Anshu},
  \citenamefont {Arad},\ and\ \citenamefont
  {Vidick}}]{anshuSimpleProofDetectability2016}%
  \BibitemOpen
  \bibfield  {author} {\bibinfo {author} {\bibfnamefont {A.}~\bibnamefont
  {Anshu}}, \bibinfo {author} {\bibfnamefont {I.}~\bibnamefont {Arad}}, \ and\
  \bibinfo {author} {\bibfnamefont {T.}~\bibnamefont {Vidick}},\ }\href
  {\doibase 10.1103/PhysRevB.93.205142} {\bibfield  {journal} {\bibinfo
  {journal} {Physical Review B}\ }\textbf {\bibinfo {volume} {93}},\ \bibinfo
  {pages} {205142} (\bibinfo {year} {2016})}\BibitemShut {NoStop}%
\bibitem [{Note7()}]{Note7}%
  \BibitemOpen
  \bibinfo {note} {We used $T_m(x)>\protect \frac {1}{2}e^{m \protect \arccosh
  x}>\protect \frac {1}{2}e^{2m \tanh (\protect \frac {1}{2}\protect \arccosh
  x)}=\protect \frac {1}{2}e^{2m\protect \sqrt {(x-1)/(x+1)}}$ for
  $x>1$.}\BibitemShut {Stop}%
\end{thebibliography}%
\appendix
\onecolumngrid
\clearpage

\section{Halperin's bound}
\label{sec:Halperin}
Here we review the argument for the previous best bound $z\geq2-\eta$ obtained by Halperin~\cite{PhysRevB.8.4437} in terms of the Rokhsar-Kivelson Hamiltonian. 

Let us consider the autocorrelation function of the total magnetization $\hat{M}\coloneqq\sum_{i\in\Lambda_L}\hat{s}_i^z$.
\begin{align}
C(t)\coloneqq \langle\Phi_0|\hat{M} e^{-\hat{H}t}(1-\hat{G})\hat{M}|\Phi_0\rangle,
\end{align}
where $\hat{G}=|\Phi_0\rangle\langle\Phi_0|$ is the projector onto the ground state of $\hat{H}=\sum_{i\in \Lambda_L}\hat{H}_i$ in Eq.~\eqref{heatbath2}. Let $|\Phi_n\rangle$ ($n=1,2,\cdots$) be excited states and $E_n$ be their eigenenergies. It follows that
\begin{align}
C(t)&=\sum_n|\langle\Phi_n|\hat{M}|\Phi_0\rangle|^2e^{-E_nt}.
\end{align}
We define the characteristic time $\bar{\tau}$ by
\begin{align}
\bar{\tau}&\coloneqq\frac{\int_0^\infty dtC(t)}{C(0)}=\frac{\langle\Phi_0|\hat{M}\frac{1-\hat{G}}{\hat{H}}\hat{M}|\Phi_0\rangle}{\langle\Phi_0|\hat{M}(1-\hat{G})\hat{M}|\Phi_0\rangle}=\frac{\sum_{n}\frac{|\langle\Phi_n|\hat{M}|\Phi_0\rangle|^2}{E_n}}{\sum_n|\langle\Phi_n|\hat{M}|\Phi_0\rangle|^2}\leq\frac{1}{\epsilon}=\tau.
\end{align}
To derive a lower bound of this quantity, let us define the uniform susceptibility ($T$ is the temperature)
\begin{align}
\chi\coloneqq\frac{1}{TL^d}\langle\Phi_0|\hat{M}(1-\hat{G})\hat{M}|\Phi_0\rangle
\end{align}
and a quantity
\begin{align}
R&\coloneqq\frac{1}{L^d}\langle\Phi_0|\hat{M}\hat{H}(1-\hat{G})\hat{M}|\Phi_0\rangle\notag\\
&=\frac{1}{2L^d}\langle\Phi_0|[\hat{M},[\hat{H},\hat{M}]]|\Phi_0\rangle\notag\\
&=\frac{1}{L^d}\sum_i\langle\Phi_0|\frac{1}{\cosh(\beta\sum_{j\in B_i}\hat{\sigma}_j^z)}\hat{\sigma}_i^x|\Phi_0\rangle.
\end{align}

The Schwartz inequality
\begin{align}
\langle\Phi_0|\hat{M}\frac{1-\hat{G}}{\hat{H}}\hat{M}|\Phi_0\rangle\langle\Phi_0|\hat{M}\hat{H}(1-\hat{G})\hat{M}|\Phi_0\rangle\geq\langle\Phi_0|\hat{M}(1-\hat{G})\hat{M}|\Phi_0\rangle^2
\end{align}
leads to
\begin{align}
\tau\geq\bar{\tau}\geq\frac{T}{R}\chi.
\end{align}
Now, we take the $L\to\infty$ limit. The susceptibility behaves as $\chi \simeq |T-T_c|^{-\gamma}$, while $R$ does not show singularity around the critical point $T=T_c$.
Hence, if we assume $\bar{\tau} \simeq |T-T_c|^{-z\nu}$, one gets 
\begin{align}
z\nu\geq\gamma.
\end{align}

\section{Derivation of the lower bound}
\label{app:LB}
Here we review Simon and Lieb's results~\cite{Boel,Simon,Lieb} on the correlation function of Ising model \eqref{isingbw}.
In this section, we specify the system size in the expectation value \eqref{exp} and write $\langle O\rangle_L$.
We recall that we always use periodic boundary conditions.
\subsection{Statement}
Suppose that $c(L)$ defined in \eqref{deflambda} satisfies $0<c(L)< 1$ for some $L=1,3,5,\cdots$. Then the correlation function $\langle\sigma_o\sigma_i\rangle_{L'}$ decays exponentially for any $L'>L$, i.e.,
\begin{align}
\langle \sigma_o\sigma_i\rangle_{L'}\leq Ce^{-\|i\|_\infty/\xi},\label{B1}
\end{align}
where $\xi\coloneqq -\frac{L+1}{2\log c(L)}>0$, $C\coloneqq c(L)^{-1}$, and $\|i\|_\infty$ is the $L^\infty$-norm defined by
\begin{align}
\|i\|_\infty\coloneqq\max_{a}\big\{\min\{|i^{(a)}|,L-|i^{(a)}|\}\big\}.
\end{align}
Note that \ref{B1} implies
\begin{align}
\limsup_{L'\to\infty}\sum_{i\in\Lambda_{L'}}\langle\sigma_o\sigma_i\rangle_{L'}<\infty
\end{align}
and hence $\beta<\beta_c$. See~\cite{FriedliVelenik,SimonBook}. We thus find that $c(L)>1$ for any $L=1,3,5,\cdots$ if $\beta\geq\beta_c$.

\subsection{Proof}
Let us define an integer $n_i$ by
\begin{align}
n_i\coloneqq\Big\lfloor\frac{2}{L+1}\|i\|_\infty\Big\rfloor,
\end{align}
where $\lfloor r\rfloor$  represents the integer part of a real number $r$. If $n_i\geq1$, the Simon-Lieb inequality~\cite{Boel,Simon,Lieb} implies 
\begin{align}
\langle \sigma_o\sigma_i\rangle_{L'}\leq \sum_{j\in\partial\Lambda_{L}}\sum_{k_1\in B_j\cap\partial\Lambda_{L+2}}\langle \sigma_o\sigma_j\rangle_{L}^{\text{free}}\langle\sigma_{k_1}\sigma_i\rangle_{L'},\label{B5}
\end{align}
where $B_i$ denotes the set of lattice sites adjacent to $i$ and $\langle \sigma_o\sigma_j\rangle_{L}^{\text{free}}$ is the correlation function with free boundary conditions, which can be bounded as $\langle \sigma_o\sigma_j\rangle_{L}^{\text{free}}\leq\langle \sigma_o\sigma_j\rangle_{L}$ by the Griffiths second inequality. Thus the right-hand side of \ref{B5} can be bounded from above to give
\begin{align}
\langle \sigma_o\sigma_i\rangle_{L'}\leq \sum_{j\in\partial\Lambda_{L}}|B_j|\langle \sigma_o\sigma_j\rangle_{L}\max_{k_1\in\partial\Lambda_{L+2}}  \langle \sigma_{k_1}\sigma_i\rangle_{L'}.
\end{align}
Plugging $|B_j|=2d$ and the definition of $c(L)$ in \eqref{deflambda}, we find
\begin{align}
\langle \sigma_o\sigma_i\rangle_{L'}\leq c(L)\max_{k_1\in\partial\Lambda_{L+2}}  \langle \sigma_{o}\sigma_{i-k_1}\rangle_{L'}.
\end{align}
Repeating this process $n_i$ times and using \eqref{correlationineq}, we find
\begin{align}
\langle \sigma_o\sigma_i\rangle_{L'}\leq c(L)^{n_i}\max_{k_1,k_2,\cdots,k_{n_i}\in\partial\Lambda_{L+2}}  \langle \sigma_{o}\sigma_{i-\sum_{j=1}^{n_i}k_j}\rangle_{L'}
\leq c(L)^{n_i}\leq Ce^{-\|i\|_\infty/\xi},
\end{align}
where we used  $\lfloor r\rfloor\geq r-1$ in the last step.

\section{Generalization}
\label{Generalization}
Our discussion presented in the main text can be readily extended to other algorithms in which $P_i(\bm{\sigma})$  (i) satisfies the detailed balance condition \eqref{detailed}, (ii) depends only on spins in a finite distance from the site $i$, (iii) gives the transition matrix $W$ with ergodicity. 

To see this, note that the corresponding Rokhsar-Kivelson Hamiltonian
\begin{align}
\hat{H}^{\text{(RK)}}=\sum_{i\in\Lambda_L}\hat{H}_i^{\text{(RK)}},
\end{align}
defined by \eqref{defWP} and \eqref{HW}, is frustration free and its unique ground state is given by $|\Phi_0\rangle$ in \eqref{GS}.

Let $|i,n\rangle$ ($n=1,2,\cdots$) be the eigenvectors of $\hat{H}_i^{\text{(RK)}}$ with nonzero eigenvalues.
Then $\hat{P}_i^{\text{(RK)}}\coloneqq\sum_{n}|i,n\rangle\langle i,n|$ is the projector version of $\hat{H}_i^{\text{(RK)}}$.  Let $\epsilon_L^{\text{(RK)}}$ and $\tilde{\epsilon}_L^{\text{(RK)}}$ be the second smallest eigenvalue of $\hat{H}^{\text{(RK)}}$ and $\sum_{i\in\Lambda_L}\hat{P}_i^{\text{(RK)}}$, respectively. Then we have
\begin{align}
\hat{H}^{\text{(RK)}}\leq \max_{i\in\Lambda_L}\|\hat{H}_i^{\text{(RK)}}\|\sum_{i\in\Lambda_L}\hat{P}_i^{\text{(RK)}},
\end{align}
which implies
\begin{align}
\epsilon_L^{\text{(RK)}}\leq \max_{i\in\Lambda_L}\|\hat{H}_i^{\text{(RK)}}\| \tilde{\epsilon}_L^{\text{(RK)}}.
\end{align}
Therefore, our discussion for projector Hamiltonians is sufficient.

For example, in the Metropolis algorithm, the flipping rate $P_i(\bm{\sigma})$ is given by
\begin{align}
P_i(\bm{\sigma})\coloneqq\min(1,e^{-2\beta\sigma_i\sum_{j\in B_i}\sigma_j})
\end{align}
and the corresponding Rokhsar-Kivelson Hamiltonian is
\begin{align}
\hat{H}_i^{\text{(RK)}}&=\min(e^{\beta\sum_{j\in B_i}\hat{\sigma}_j^z},e^{-\beta\sum_{j\in B_i}\hat{\sigma}_j^z})\Big(e^{-\beta\hat{\sigma}_i^z{}\sum_{j\in B_i}\hat{\sigma}_j^z}-\hat{\sigma}_i^x\Big).
\end{align}
The largest eigenvalue of $\hat{H}_i^{\text{(RK)}}$ gives $\|\hat{H}_i^{\text{(RK)}}\|=2$.

Another simple choice will be
\begin{align}
P_i(\bm{\sigma})\coloneqq e^{-\beta\sigma_i\sum_{j\in B_i}\sigma_j},
\end{align}
which corresponds to  
\begin{align}
\hat{H}_i^{\text{(RK)}}&=e^{-\beta\hat{\sigma}_i^z{}\sum_{j\in B_i}\hat{\sigma}_j^z}-\hat{\sigma}_i^x
\end{align}
with $\|\hat{H}_i^{\text{(RK)}}\|=2\cosh(2d\beta)$.

\section{Interaction graph}
\label{app:IG}
Let us define some notions that will play fundamental roles in Appendix~\ref{app:DL} and \ref{app:GH}.

Consider a quantum system with Hilbert space $\mathcal{H}$ and let $\hat{H}_i$ with $i\in V$ be an arbitrary collection of operators on $\mathcal{H}$.  Here, $V$ is an arbitrary finite set, which we identify as the set of vertices. We then define the set of edges $E$ as a collection of unordered pairs $\{i,j\}$ such that $[\hat{H}_i,\hat{H}_j]\neq0$. The graph $(V,E)$ is called the interaction graph associated with the collection of $\hat{H}_i$

Let $V_i\coloneqq \big\{j\in V\mid[\hat{H}_i,\hat{H}_j]\neq0\big\}$ be the set of vertices adjacent to $i\in V$.
Let $g\coloneqq\max_{i\in V}|V_i|$ be the maximum degree and $c$ be the chromatic number of the interaction graph. We assume $c\geq2$. Also, let $d(i,j)$ be the graph theoretic distance (the shortest-path distance) between $i\in V$ and $j\in V$ on the interaction graph.

\section{Detectability Lemma}
\label{app:DL}
The proof of the Gosset--Huang inequality is based on the detectability lemma~\cite{aharonovDetectabilityLemmaIts2011,anshuSimpleProofDetectability2016}.

\subsection{Statement}
Suppose that the Hamiltonian
\begin{align}
\hat{H} = \sum_{i=1}^N\hat{H}_i
\end{align}
is a sum of projectors $\hat{H}_i^2=\hat{H}_i$ and is frustration free, i.e., all $\hat{H}_i$'s can be simultaneously minimized. We assume that the lowest eigenvalue of $\hat{H}$ is 0.

Let $\hat{G}$ be the projector onto the space of all the ground states of $\hat{H}$. Note that $\hat{H}_i\hat{G}=0$ for any $i$. Let $\epsilon$ be the smallest eigenvalue of $\hat{H}$ other than $0$. Let us write $\hat{P}_i \coloneqq 1 - \hat{H}_i$ and define
\begin{align}
\hat{P} \coloneqq \hat{P}_{\sigma(1)}\hat{P}_{\sigma(2)}\cdots\hat{P}_{\sigma(N)},
\end{align}
where $\sigma$ is an arbitrary permutation of $1,2,\cdots,N$. Furthermore, we consider the interaction graph $(V,E)$ associated with $\hat{H}_i$ and $V=\{1,2,\cdots,N\}$ and define the maximal degree $g$ (see Sec.~\ref{app:IG}). Then,
\begin{align}
\|\hat{P}-\hat G\| \leq \sqrt{\frac{g^2}{g^2+\epsilon}}.\label{DL}
\end{align}

\subsection{Proof}
Take a state $|\psi\rangle$ and we perform the following operations to the quantity $\|\hat{H}_i\hat{P}_{\sigma(j)}\cdots \hat{P}_{\sigma(N)}|\psi\rangle\|$.
If $\hat{H}_i$ commutes with $\hat{H}_{\sigma(j)}$,
\begin{align}
\|\hat{H}_i\hat{P}_{\sigma(j)}\cdots \hat{P}_{\sigma(N)}|\psi\rangle\| &
=\|\hat{P}_{\sigma(j)}\hat{H}_i\hat{P}_{\sigma(j+1)}\cdots \hat{P}_{\sigma(N)}|\psi\rangle\|\nonumber\\
&\leq \|\hat{H}_i\hat{P}_{\sigma(j+1)}\cdots \hat{P}_{\sigma(N)}|\psi\rangle\|.
\end{align}
Otherwise, 
\begin{align}
\| \hat{H}_i\hat{P}_{\sigma(j)}\cdots \hat{P}_{\sigma(N)}|\psi\rangle\|&=\| \hat{H}_i(1-\hat{H}_{\sigma(j)})\hat{P}_{\sigma(j+1)}\cdots \hat{P}_{\sigma(N)}|\psi\rangle\|\nonumber\\
&\leq \| \hat{H}_i\hat{P}_{\sigma(j+1)}\cdots \hat{P}_{\sigma(N)}|\psi\rangle\|+\|\hat{H}_i\hat{H}_{\sigma(j)}\hat{P}_{\sigma(j+1)}\cdots \hat{P}_{\sigma(N)}|\psi\rangle\| \nonumber\\
&\leq \|\hat{H}_i\hat{P}_{\sigma(j+1)}\cdots \hat{P}_{\sigma(N)}|\psi\rangle\|+\|\hat{H}_{\sigma(j)}\hat{P}_{\sigma(j+1)}\cdots \hat{P}_{\sigma(N)}|\psi\rangle\|.
\end{align}
Note that the first term in the last expression is the original quantity but $j$ is shifted to $j+1$. 
If we start from $\|\hat{H}_i\hat{P}|\psi\rangle\|$ and repeat this procedure, $\hat{H}_i$ never reaches $|\psi\rangle$, because $\hat{H}_i\hat{P}_i=0$. At the end, we obtain the following terms:
\begin{align}
\|\hat{H}_i\hat{P}|\psi\rangle\|
\leq \sum_{\sigma(l)\in V_i} \|\hat{H}_{\sigma(l)}\hat{P}_{\sigma(l+1)}\cdots\hat{P}_{\sigma(N)}|\psi\rangle\|.
\end{align}
Since the square of an average is less than or equal to the average of the squares,
\begin{align}
\| \hat{H}_i\hat{P}|\psi\rangle\|^2&
\leq |V_i|^2\Big(\frac1{|V_i|} \sum_{\sigma(l)\in V_i} \|\hat{H}_{\sigma(l)} \hat{P}_{\sigma(l+1)} \cdots \hat{P}_{\sigma(N)} |\psi\rangle\|\Big)^2 \nonumber\\
&
\leq |V_i|^2\Big(\frac{1}{|V_i|} \sum_{\sigma(l)\in V_i} \|\hat{H}_{\sigma(l)} \hat{P}_{\sigma(l+1)} \cdots \hat{P}_{\sigma(N)}|\psi\rangle \|^2\Big)\nonumber\\
&\leq  g\sum_{\sigma(l)\in V_i} \|\hat{H}_{\sigma(l)} \hat{P}_{\sigma(l+1)} \cdots \hat{P}_{\sigma(N)}|\psi\rangle \|^2.
\end{align}
Therefore,
\begin{align}
\langle\psi|\hat{P}^\dagger\hat{H}\hat{P}|\psi\rangle&= \sum_{i=1}^{N}\langle\psi|\hat{P}^\dagger\hat{H}_i\hat{P}|\psi\rangle= \sum_{i=1}^{N}\langle\psi|\hat{P}^\dagger\hat{H}_i^2\hat{P}|\psi\rangle =\sum_{i=1}^{N} \| \hat{H}_i\hat{P}|\psi\rangle\|^2\nonumber\\
&\leq g\sum_{i=1}^{N}   \sum_{\sigma(l)\in V_i} \|\hat{H}_{\sigma(l)} \hat{P}_{\sigma(l+1)} \cdots \hat{P}_{\sigma(N)}|\psi\rangle \|^2.
\end{align}
Exchanging the order of the summations, we find
\begin{align}
\langle\psi|\hat{P}^\dagger\hat{H}\hat{P}|\psi\rangle&\leq g^2 \sum_{l=1}^{N} \|\hat{H}_{\sigma(l)}\hat{P}_{\sigma(l+1)} \cdots \hat{P}_{\sigma(N)}|\psi\rangle \|^2 \nonumber\\
&=g^2 \sum_{l=1}^{N} \|(1-\hat{P}_{\sigma(l)})\hat{P}_{\sigma(l+1)} \cdots \hat{P}_{\sigma(N)}|\psi\rangle \|^2 \nonumber\\
&=g^2\sum_{l=1}^{N} \big(\|\hat{P}_{\sigma(l+1)} \cdots \hat{P}_{\sigma(N)}|\psi\rangle\|^2 -\|\hat{P}_{\sigma(l)} \cdots \hat{P}_{\sigma(N)}|\psi\rangle\|^2 \big)\nonumber\\
& = g^2\big(\||\psi\rangle\|^2 -\|\hat{P}|\psi\rangle\|^2 \big).
\end{align}
Applying this result to $|\psi^\perp\rangle \coloneqq (1-\hat G)|\psi\rangle$ and using the definition of $\epsilon$, 
\begin{align}
\epsilon \| \hat{P}|\psi^\perp\rangle\|^2 &\leq \langle\psi^\perp|\hat{P}^\dagger \hat{H}\hat{P}|\psi^\perp\rangle \leq g^2(\||\psi^\perp\rangle\|^2-\| \hat{P}|\psi^\perp\rangle\|^2),
\end{align}
which can be rewritten as
\begin{align}
\| \hat{P}|\psi^\perp\rangle\|\leq \sqrt{\frac{g^2}{g^2+\epsilon}}\||\psi^\perp\rangle\|.
\end{align}
Therefore, using $\hat{P}\hat{G}=\hat{G}$,
\begin{align}
\|(\hat{P}-\hat G)|\psi\rangle\|=\| \hat{P}(1-\hat G)|\psi\rangle\|= \| \hat{P}|\psi^\perp\rangle\|\leq \sqrt{\frac{g^2}{g^2+\epsilon}}\||\psi^\perp\rangle\|\leq\sqrt{\frac{g^2}{g^2+\epsilon}}\||\psi\rangle\|
\end{align}
for any $|\psi\rangle$, which implies \eqref{DL}.

\section{Gosset--Huang inequality}
\label{app:GH}
\subsection{Statement}
Let $V$ be a finite set and suppose that the Hamiltonian
\begin{align}
\hat{H} = \sum_{i\in V}\hat{H}_i
\end{align}
is a sum of projectors $\hat{H}_i^2=\hat{H}_i$ and is frustration free.
We assume that the lowest eigenvalue of $\hat{H}$ is 0.
Let $\hat{G}$ be the projector onto the space of all the ground states of $\hat{H}$. Let $\epsilon$ be the smallest eigenvalue of $\hat{H}$ other than $0$. 

We consider the interaction graph $(V,E)$ associated with $\hat{H}_i$ (see Sec.~\ref{app:IG}). Let $g$ be the maximum degree and $c$ be the chromatic number of the interaction graph. We assume $c\geq2$. Also, let $d(i,j)$ be the graph theoretic distance between $i,j\in V$ on the interaction graph. 

Then, for any operators $\hat{O}$ and $\hat{O}'$, the following inequality holds~\cite{PhysRevLett.116.097202}:
\begin{align}
\frac{\langle\Phi_0|\hat{O}(1-\hat G)\hat{O}'|\Phi_0\rangle}{\|\hat{O}^\dagger|\Phi_0\rangle\| \|\hat{O}'|\Phi_0\rangle\|}\leq 2e^2 \exp\Big(-\frac{D(\hat{O},\hat{O}')-1}{c-1}\sqrt{\frac{\epsilon}{g^2+\epsilon}}\Big), \label{GHa}
\end{align}
where $\epsilon$ is the smallest eigenvalue of $\hat{H}$ apart from $0$ and 
\begin{align}
D(\hat{O},\hat{O}')\coloneqq \min_{i,j}\big\{d(i,j)\,\big|\,[\hat{H}_i,\hat{O}]\neq0,[\hat{H}_j,\hat{O}']\neq0\big\}.\label{defD}
\end{align}

\subsection{Proof}
We assume a vertex coloring of the interaction graph and decompose the set $V$ according to the colors of vertices as $V=\bigcup_{j=1}^cV^{(j)}$ so that $\{i,i'\}\notin E$ for any $i,i'\in V^{(j)}$. We shall relabel the operators $\hat{H}_i$ as $\hat{H}_i^{(j)}$ where $j\in\{1,\cdots,c\}$ and $i\in V^{(j)}$.
By definition, $\hat{H}_i^{(j)}$ with the same color $j$ commutes with each other.
Set $\hat{P}_i^{(j)} \coloneqq 1 - \hat{H}_i^{(j)}$ and 
$\hat{P}^{(j)}\coloneqq \prod_{i\in V^{(j)}}\hat{P}_i^{(j)}$ for each $j\in\{1,\cdots,c\}$.
Furthermore, define
\begin{align}
\hat{P} \coloneqq \hat{P}^{(c)}\cdots\hat{P}^{(2)}\hat{P}^{(1)}.
\end{align}

Let $n$ be a nonnegative integer. Note that $(\hat P^\dagger\hat P)^n$ contains $2n(c-1)+1$ layers.
The definition of $D(\hat{O},\hat{O}')$ in \eqref{defD} implies that any path that connects $\hat{O}$ and $\hat{O}'$ on the interaction graph contains $D(\hat{O},\hat{O}')+1$ non-commuting operators. 
However, as far as 
\begin{align}
n\leq m\coloneqq\Big\lfloor\frac{D(\hat{O},\hat{O}')-1}{2(c-1)}\Big\rfloor,
\end{align}
$2n(c-1)+1\leq D(\hat{O},\hat{O}')$. Hence, using $\hat{P}_i^{(j)}|\Phi_0\rangle=|\Phi_0\rangle$ for any $i,j$, we obtain
\begin{align}
\langle\Phi_0|\hat{O}(\hat P^\dagger\hat P)^n\hat{O}'|\Phi_0\rangle=\langle\Phi_0|\hat{O}\hat{O}'|\Phi_0\rangle,
\end{align}
For any $m$-th order polynomial $Q_m(x)$ with $Q_m(1)=1$, we have
\begin{align}
\langle\Phi_0|\hat{O}\hat{O}'|\Phi_0\rangle=\langle\Phi_0|\hat{O} Q_m(\hat P^\dagger \hat P)\hat{O}'|\Phi_0\rangle.
\end{align}
Writing $\hat G^\perp \coloneqq 1-\hat G$ and using $\hat{P}\hat{G}=\hat{G}$,
\begin{align}
(\hat P^\dagger\hat P)^n - \hat G = (\hat P^\dagger\hat P-\hat G)^n = (\hat P^\dagger \hat P-\hat G)^n \hat G^\perp
\end{align}
for $n\geq1$. For $n=0$,
\begin{align}
(\hat P^\dagger \hat P)^0-\hat G = \hat G^\perp=(\hat P^\dagger \hat P-\hat G)^0 \hat G^\perp
\end{align}
Therefore,
\begin{align}
\langle\Phi_0|\hat{O}(1-\hat G)\hat{O}'|\Phi_0\rangle
&
= \langle\Phi_0|\hat{O}(Q_m(\hat P^\dagger\hat P)-\hat G)\hat{O}' |\rangle \nonumber\\
&
= \langle\Phi_0|\hat{O} Q_m(\hat P^\dagger\hat P-\hat G)\hat G^\perp\hat{O}'|\Phi_0\rangle \nonumber\\
&
\leq \|\hat{O}^\dagger|\Phi_0\rangle \| \|\hat{O}'|\Phi_0\rangle \| \| Q_m(\hat P^\dagger\hat P-\hat G)\|.
\label{bound of correlation function}
\end{align}

Now we set $\delta \coloneqq \epsilon/(g^2+\epsilon)$. 
Since the detectability lemma \eqref{DL} implies $\|\hat P^\dagger\hat P -\hat G\| \leq 1-\delta$, we obtain
\begin{align}
\| Q_m(\hat P^\dagger\hat P-\hat G)\|
\leq \max_{0 \leq x \leq 1-\delta} |Q_m(x)|.
\label{mini max problem}
\end{align}
We set
\begin{align}
Q_m(x) = \frac{T_m(\frac{2x}{1-\delta}-1)}{T_m(\frac{2}{1-\delta}-1)}.
\end{align}
Here, $T_m(x)$ is the $m$-th order Chebyshev polynomial of the fist kind defined by $T_m(x)=\cos(m \arccos x)$ ($|x|\leq1$) and $T_m(x)=\cosh(m \arccosh x)$  ($x>1$), which satisfies
\begin{align}
T_m(x)> \frac{1}{2}e^{2m\sqrt{(x-1)/(x+1)}}
\end{align}
for $x >1$ \footnote{We used $T_m(x)>\frac{1}{2}e^{m \arccosh x}>\frac{1}{2}e^{2m \tanh(\frac{1}{2}\arccosh x)}=\frac{1}{2}e^{2m\sqrt{(x-1)/(x+1)}}$ for $x>1$.} and $|T_m(x)| \leq 1$ for $|x| \leq 1$. Hence, $Q_m(x)$ satisfies
\begin{align}
|Q_m(x)|&\leq\frac{1}{T_m(\frac{1+\delta}{1-\delta})} \leq 2{{e}}^{-2m\sqrt\delta} = 2\exp\Big(-2m\sqrt{\frac{\epsilon}{g^2+\epsilon}}\Big)
\label{bound of Q_m(x)}
\end{align}
for $0 \leq x \leq 1-\delta$. Therefore, using $\lfloor r\rfloor\geq r-1$, 
\begin{align}
\frac{\langle\Phi_0|\hat{O}(1-\hat G)\hat{O}'|\Phi_0\rangle}{\|\hat{O}^\dagger |\Phi_0\rangle\| \|\hat{O}'|\Phi_0\rangle\|}
&\leq 2\exp\Big(-2m\sqrt{\frac{\epsilon}{g^2+\epsilon}}\Big)\notag\\
&\leq 2\exp\Big(2\sqrt{\frac{\epsilon}{g^2+\epsilon}}\Big)\exp\Big(-\frac{D(\hat{O},\hat{O}')-1}{c-1}\sqrt{\frac{\epsilon}{g^2+\epsilon}}\Big).
\end{align}
Finally using $\sqrt{\epsilon/(g^2+\epsilon)}\leq1$, we arrive at \eqref{GHa}.

\end{document}